\RequirePackage{fix-cm}
\documentclass[smallextended]{svjour3}       
\smartqed  
\usepackage{amsmath}
\usepackage{bm}
\usepackage{multirow}

\usepackage{graphicx}
\begin{document}

\title{Comment on "Quantum teleportation and information splitting via four-qubit cluster state and a Bell state"%
\thanks{Project supported by the National Key Research and Development Program of China (Grant No.2017YFB0802000), National Natural Science Foundation of China (Grant No.61602378,61771877,61502376), and Natural Science Basic Research Plan in Shanxi Province of China(Grant No.2016JQ6033).}
}


\author{Zhang MeiLing  \and Shi Sha \and Liu YuanHua  \and Zheng QingJi \and Wang YunJiang
}


\institute{M. Zhang, Y. Liu, Q. Zheng, \at
              School of Communication and Information Engineering,
              XI'AN University of Posts \& Telecommunications, Xi'an 710121, China. \\
          S. Shi, Y. Wang\at
              Xidian University, Xian 710071, China. \\
              \email{zhangmlwy@126.com}
}

\date{Received: date / Accepted: date}

\maketitle

\begin{abstract}
We study the quantum controlled and probabilistic teleportation protocol via a four-cluster state (Front. Phys. (2017) 12: 120306). The protocol cannot achieve the goal that if the teleportation fails, it can be repeated without copies of the teleported state. And the reason is that all the information of the teleported state shifts to the receiver's particle after Bell measurement and the information cannot shift back to the sender without other entangled states. To realize the claimed goal,  a new quantum controlled and probabilistic teleportation protocol using the same four-cluster state is presented.
Because no information of the teleported stated is lost during the whole teleportation process,  the teleportation process can be repeated until the teleportation success.
\end{abstract}
\section{Review and Analysis of Ram\'irez's controlled and probabilistic teleportation protocol \cite{Front. Phys. (2017) 12: 120306}}
In \cite{Front. Phys. (2017) 12: 120306}, Ram\'{i}rez et. al. proposed a  quantum probabilistic teleportation of a qubit. They claimed that if the teleportation fails, the teleportation process can be repeated without  copies of the teleported state, provided that the three participants Alice, Bob, and Chika share another pair of entangled qubits. The detail teleportation process is described briefly as follows. \par
Suppose that an arbitrary single qubit state that Alice wishes to teleport to Chika under the control of Bob is $|\zeta\rangle_A = a|0\rangle + b|1\rangle$, where $|a|^2 +|b|^2 =1$.
All participants share a quantum channel cluster state $|\varphi\rangle_{1234} = \alpha |0000\rangle +\beta |1010\rangle + \gamma |0101\rangle - \eta|1111\rangle$ with $|\alpha|^2 + |\beta|^2 + |\gamma|^2 + |\eta|^2 = 1 $, where particle $1$ belongs to Alice, $2$ and $3$ belong to Bob and $4$ to Chika. Then the whole system can be expressed as $|\phi\rangle_{A1234} = |\zeta\rangle_A|\varphi\rangle_{1234}$.
\begin{description}
\item{\bf{Step 1 }}
 Alice performs Bell measurement on her particles ($A$,1) and publishes her results using a classical channel. \par
\item{\bf{Step 2 }} If Bob agrees to assist Chika to recover the original state of the qubit $|\zeta\rangle$, he performs a Bell measurement on his particles $(2,3)$ and publishes his measurement results.\par
\item{\bf{Step 3 }} Consequently, Chika's state collapses to the corresponding state as shown in Table.~\ref{tab1} by omitting the global phase. For example, if Alice's and Bob's BM results are $|\psi^+\rangle$ and $|\phi^-\rangle$, respectively, Chika's state collapses to $b \alpha|0\rangle + a\eta|1\rangle $ .
In order to make the state of qubit $4$ be the form of $|\varphi\rangle_4 = \delta_0 a|0\rangle + \delta_1 b |1\rangle$,
proper Pauli operators $\sigma_I$ or $\sigma_X$ are needed to apply on the qubit $4$. As the example above illustrates, a Pauli bit-flip operator $\sigma_X$ is applied to change the state of qubit $4$ to $|\varphi\rangle_4 = a\eta|0\rangle + b \alpha|1\rangle $. \par
\begin{table}[!htbp]
\caption{Alice and Bob's BSM results and the corresponding result of Chika's state}
\centering
\begin{tabular}{|c|c|c|c|c|c|}
\hline
\multicolumn{2}{|c|}{ \multirow{2}*{Chika's state} }& \multicolumn{4}{c|}{Bob's BM result} \\
\cline{3-6}
\multicolumn{2}{|c|}{} & $ |\phi^+\rangle$ & $ |\phi^-\rangle$ & $|\psi^+\rangle $ & $|\psi^-\rangle $\\
\hline
Alice's        &    $ |\phi^+\rangle$  & $a \alpha|0\rangle - b\eta|1\rangle $  & $ a \alpha|0\rangle + b\eta|1\rangle $
                                     & $b \beta|0\rangle + a\gamma|1\rangle $ & $ b \beta|0\rangle - a\gamma|1\rangle $\\
\cline{2-6}
BM           & $ |\phi^-\rangle$     & $a \alpha|0\rangle + b\eta|1\rangle $  & $a \alpha|0\rangle - b\eta|1\rangle $
                                     & $b \beta|0\rangle - a\gamma|1\rangle $ & $b \beta|0\rangle + a\gamma|1\rangle $\\
\cline{2-6}
result      &$|\psi^+\rangle $       & $b \alpha|0\rangle - a\eta|1\rangle $  & $ b \alpha|0\rangle + a\eta|1\rangle $
                                     & $a \beta|0\rangle + b\gamma|1\rangle $ & $ a \beta|0\rangle - b\gamma|1\rangle $ \\
\cline{2-6}
            &$|\psi^-\rangle $       & $b \alpha|0\rangle + a\eta|1\rangle $  & $ b \alpha|0\rangle - a\eta|1\rangle $
                                     & $a \beta|0\rangle - b\gamma|1\rangle $ & $ a \beta|0\rangle + b\gamma|1\rangle $\\
\hline
\end{tabular}
\label{tab1}
\end{table}
\item{\bf{Step 4 }} Chika introduces an auxiliary qubit $e$ with state $|0\rangle_e$, then $|\varphi\rangle_4|0\rangle_e =\delta_0 a|00\rangle_{4e} + \delta_1 b |10\rangle_{4e}$. Next, she performs a C-NOT operation on qubit pair $(4,e)$ to obtain
 \begin{equation}
\begin{aligned}
&|\varphi'\rangle_{4e} = \delta_0 a|00\rangle_{4e} + \delta_1 b |11\rangle_{4e} \\
                      &= [(a|0\rangle +  b |1\rangle)_{4}(\delta_0 |0\rangle + \delta_1  |1\rangle)_{e}
                         +(a|0\rangle -  b |1\rangle)_{4}(\delta_0 |0\rangle - \delta_1  |1\rangle)_{e}]/2
\end{aligned}
\end{equation}
\item{\bf{Step 5 }} Chika performs POVM on the auxiliary qubit $e$ by using the following operators
 \begin{equation}
 \Lambda_1 = \frac{1}{\varrho}|M_1\rangle \langle M_1|, \ \ \ \Lambda_2 = \frac{1}{\varrho}|M_2\rangle \langle M_2|, \ \ \
 \Lambda_3 = I - \Lambda_1 - \Lambda_2
 \end{equation}
 where
  \begin{equation}
 M_1 = \frac{1}{\sqrt{\varsigma}}\Big( \frac{1}{\delta_0}|0\rangle + \frac{1}{\delta_1} |1\rangle\Big), \ \
  M_2 = \frac{1}{\sqrt{\varsigma}}\Big( \frac{1}{\delta_0}|0\rangle - \frac{1}{\delta_1} |1\rangle\Big)   \ \ and \ \ \varsigma = \frac{1}{\delta_0 ^2} + \frac{1}{\delta_1 ^2}
  \end{equation}
  $I$ denotes an identity operator and  $ \varrho $ is a parameter which lies within 1 and 2 to make sure that $\Lambda_3$ is a non-negative operator, i.e. $ 1 \le \varrho \le 2$. There are three different POVM results of qubit $e$ as follows. \par
  (I) $\Lambda_1$, with a probability of $\langle \varphi'|\Lambda_1 | \varphi'\rangle = \frac{1}{4 \varrho \varsigma}$. The state of qubit $4$ collapse to the original state $a|0\rangle + b|1\rangle$. \par
  (II) $\Lambda_2$, with a probability of $\langle \varphi'|\Lambda_2 | \varphi'\rangle = \frac{1}{4 \varrho \varsigma}$. The state of qubit $4$ collapse to $a|0\rangle - b|1\rangle$, which is transformed to the original state by applying a Pauli phase-flip operator $\sigma_Z$. \par
  (III) $\Lambda_3$, with a probability of $1-\frac{1}{2 \varrho \varsigma}$. In this case, the scheme fails because Chika has not enough information to identify the state of qubit $e$. \par
  \end{description}
 It's clear that whatever results of Chika's measurement are, once Alice performs Bell measurement on qubit pair $(A, 1)$, the states of Alice's only qubit pair $(A,1)$  are in mixed states
 $\{(p_1,|\phi^{+}\rangle ), (p_1,|\phi^{-}\rangle ), (p_2,|\psi^{+}\rangle), (p_2,|\psi^{-}\rangle )\}$, where $p_1 = (|a\alpha|^2 + |a \gamma|^2 + |b \beta|^2 + |b\eta|^2)/2$ and $p_2 = (|b \alpha|^2 + |b\gamma|^2 + |a\beta|^2 + |a\eta|^2)/2 $. The state can also be represented by the density matrix
 \begin{equation}
 \rho_{A,1} = diag(p_1, p_1, p_2, p_2)
 \end{equation}
which gives no information about the right state $|\zeta\rangle$. Therefore, in this scheme, it's impossible for the state $|\zeta\rangle$ to be shifted back to qubit $A$.

\section{The proposed controlled and probabilistic teleportation protocol}
\begin{figure}
  \centering
  \centering
\includegraphics[height=5cm,width=7.5cm]{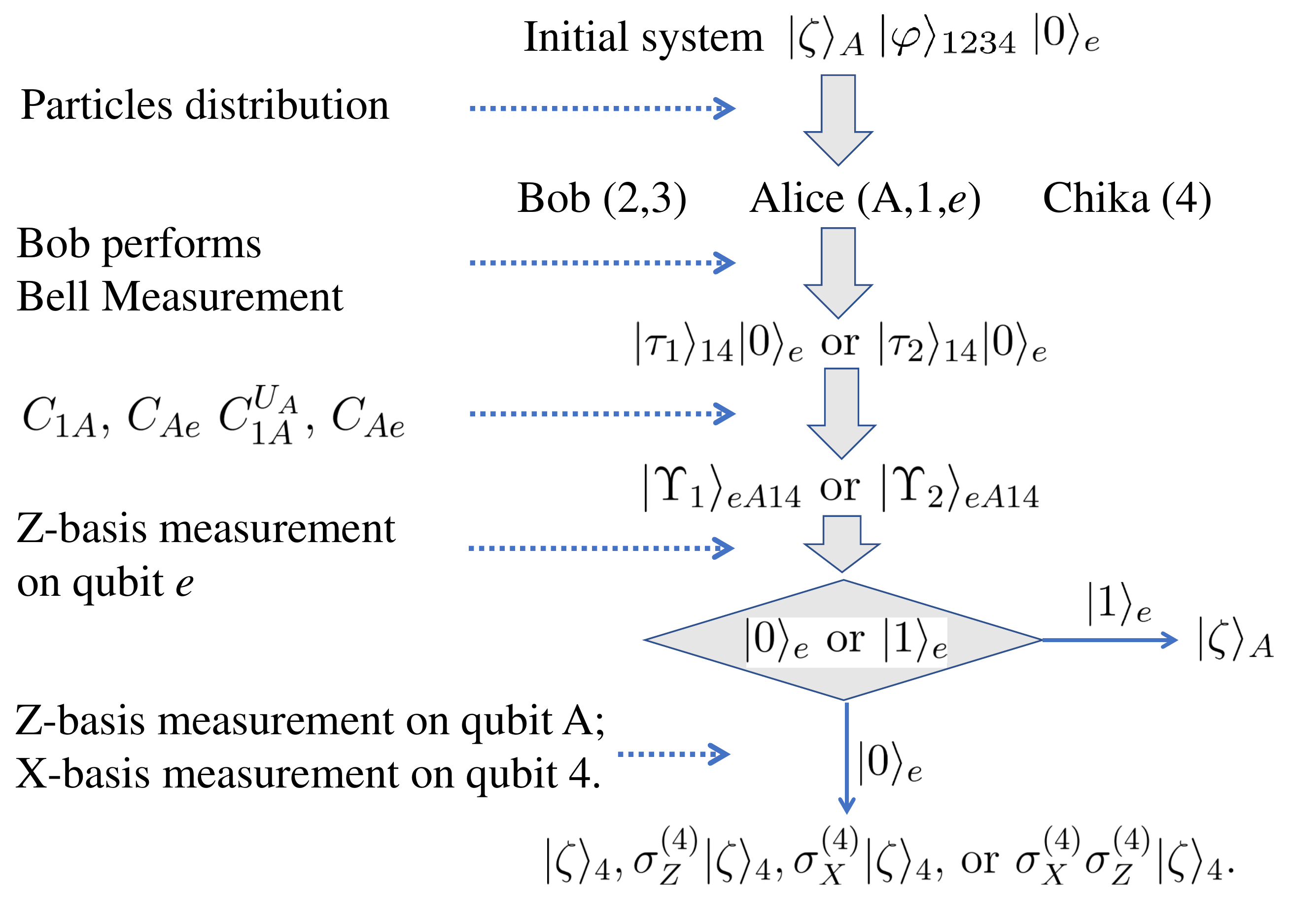}
\caption{The teleportation process in the proposed protocol}
\label{fig1}
\end{figure}

In this section, a controlled and probabilistic teleportation scheme is proposed to overcome the problem that if the teleportation process fails, the process cannot be repeated without a copy of  $|\zeta\rangle$.
 In our protocol, we change the decision maker, i.e. Alice's measurement results decide whether Chika can recover $|\zeta\rangle$ instead of Chika himself.
Here, the design idea of Roa's probabilistic teleportation without loss of information \cite{Roa2015} is adopted.
The teleported state $|\zeta\rangle_A$, quantum channel $|\varphi\rangle_{1234}$ and corresponding particles allocation are the same as Ram\'irez's scheme\cite{Front. Phys. (2017) 12: 120306}.
Without loss of generality, we can assume $\alpha$ and $\beta$ have the same phases with $\eta$ and $\gamma$, respectively, and $|\alpha| \le |\eta|$, $|\beta| \le |\gamma|$.
Next, the teleportation process is detailedly illustrated and briefly shown in Fig.~\ref{fig1}
\begin{description}
\item{\bf{Step 1 }}
 Suppose Bob agrees to assist Alice to teleport the state $|\psi\rangle_A$, he performs a Bell measurement on his particles $2$ and $3$,
with the result that the states of particle $1$ and $4$ collapse to partially entangled states shown in Table.~\ref{tab2}.
Therefore, the $|\varphi\rangle_{14}$ is either in the form of $ |\tau_1\rangle = u|00\rangle + v|11\rangle$, or in the form of $|\tau_2\rangle = u|01\rangle + v|10\rangle$. \par
\begin{table}[htbp]
 \caption{The BSM results on qubit $2$ and $3$ and their corresponding results}
 \centering
 \begin{tabular}{c | cccc}
   \hline
    $BSM_{23}$  & $| \phi^+ \rangle$
                & $| \phi^-\rangle$
                & $| \psi^+ \rangle$
                & $| \psi^- \rangle$ \\
   \hline
    $Pr_i$      & $\frac{1}{2} \big[ |\alpha |^2 + |\eta |^2 \big]$
                & $\frac{1}{2} \big[ |\alpha |^2 + |\eta |^2 \big]$
                & $\frac{1}{2} \big[ |\beta |^2 + |\gamma |^2 \big]$
                & $\frac{1}{2} \big[ |\beta |^2 + |\gamma |^2 \big]$ \\
   \hline
   $| \varphi\rangle_{14}$
                       &$\frac{\alpha|00\rangle - \eta |11\rangle}{\sqrt{|\alpha|^2+|\eta|^2}}$
                         & $\frac{\alpha|00\rangle + \eta |11\rangle}{\sqrt{|\alpha|^2+|\eta|^2}}$
                      & $\frac{ \beta|01\rangle + \gamma|10\rangle}{\sqrt{|\beta|^2+|\gamma|^2}}$
                        & $\frac{ \beta|01\rangle - \gamma  |10\rangle}{\sqrt{|\beta|^2+|\gamma|^2}}$ \\
   \hline
   $| \varphi\rangle_{14}$: $u$
                             & $\frac{\alpha  }{\sqrt{|\alpha|^2+|\eta|^2}}$
                             & $\frac{\alpha  }{\sqrt{|\alpha|^2+|\eta|^2}}$
                             & $\frac{ \beta  }{\sqrt{|\beta|^2+|\gamma|^2}}$
                             & $\frac{ \beta }{\sqrt{|\beta|^2+|\gamma|^2}}$ \\
   \hline
   $| \varphi\rangle_{14}$: $v$
                             & $\frac{ - \eta  }{\sqrt{|\alpha|^2+|\eta|^2}}$
                             & $\frac{   \eta  }{\sqrt{|\alpha|^2+|\eta|^2}}$
                             & $\frac{ \gamma  }{\sqrt{|\beta|^2+|\gamma|^2}}$
                             & $\frac{ -\gamma }{\sqrt{|\beta|^2+|\gamma|^2}}$\\
   \hline
  $\bm{\hat{n}}$  & \multicolumn{4}{c} {$\Big( \sqrt{1-\frac{u^2}{v^2}}, 0, \frac{u}{v}  \Big)$} \\
   \hline
   $Pr_{suc}$     &  \multicolumn{4}{c} {$2(|\alpha|^2+|\beta|^2)$}  \\
   \hline
    $Pr_{fail}$   & \multicolumn{4}{c} {$|\gamma|^2+|\eta|^2 - |\alpha|^2 - |\beta|^2$} \\
   \hline
 \end{tabular}
  \label{tab2}
 \end{table}
\item{\bf{Step 2 }}
Bob tells Alice the results of his measurement using a classical channel, and then Alice knows the collapsed quantum channel shared with Chika.
The state of the tripartite system $(A,1,4)$ can be expressed as
\begin{equation}
\begin{aligned}
|\zeta\rangle_A |\tau_1\rangle
=& \frac{1}{2} \big(  |\hat{\phi}^{+} \rangle_{A1}       |\zeta\rangle_4
                    +|\hat{\phi}^{-} \rangle_{A1} \sigma_Z^{(4)}   |\zeta\rangle_4  \\ &
                    +|\hat{\psi}^{+} \rangle_{A1} \sigma_X^{(4)}    |\zeta\rangle_4
                    +|\hat{\psi}^{-} \rangle_{A1} \sigma_X^{(4)} \sigma_Z^{(4)}|\zeta\rangle_4  \big)
\end{aligned}
\label{eq5}
\end{equation}
or
\begin{equation}
\begin{aligned}
|\zeta\rangle_A |\tau_2\rangle
=& \frac{1}{2} \big(  |\hat{\phi}^{+} \rangle_{A1} \sigma_X^{(4)}   |\zeta\rangle_4
                    +|\hat{\phi}^{-} \rangle_{A1} \sigma_X^{(4)}\sigma_Z^{(4)} |\zeta\rangle_4 \\ &
                    +|\hat{\psi}^{+} \rangle_{A1}             |\zeta\rangle_4
                    +|\hat{\psi}^{-} \rangle_{A1} \sigma_Z^{(4)}   |\zeta\rangle_4  \big)
\end{aligned}
\label{eq6}
\end{equation}
with $|\hat{\phi}^{\pm}\rangle = u|00\rangle \pm v|11\rangle $ and
      $|\hat{\psi}^{\pm}\rangle = v|01\rangle \pm u|10\rangle $. \par
For simplicity, Eq.~\ref{eq5} and Eq.~\ref{eq6} can be rewrote as the following unified form
\begin{equation}
|\zeta\rangle_A |\tau\rangle
= \frac{1}{2} \big(  |\hat{\phi}^{+} \rangle_{A1} |\varepsilon_1\rangle_4
                    +|\hat{\phi}^{-} \rangle_{A1} |\varepsilon_2\rangle_4
                    +|\hat{\psi}^{+} \rangle_{A1} |\varepsilon_3\rangle_4
                    +|\hat{\psi}^{-} \rangle_{A1} |\varepsilon_4\rangle_4  \big)
\end{equation}
where $|\varepsilon_1\rangle, |\varepsilon_2\rangle, |\varepsilon_3\rangle$ and  $|\varepsilon_4\rangle $ $\in \{|\zeta\rangle, \sigma_Z|\zeta\rangle, \sigma_X|\zeta\rangle, \sigma_X\sigma_Z|\zeta\rangle\}$. \par
\item{\bf{Step 3 }}
In order to distinguish four successful teleportation cases  $|\varepsilon_1\rangle, |\varepsilon_2\rangle, |\varepsilon_3\rangle$ ,  $|\varepsilon_4\rangle $ and a fail teleportation case, at least five orthogonal states that loading at least three qubits are needed to be constructed. Therefore, besides particles $A$ and $1$, another qubit should be introduced. \par
\subitem{\bf{3.1)}} The $C_{1A}$ Controlled-NOT is applied onto the the bipartite system $1A$, so the state (7) becomes
\begin{equation}
\begin{aligned}
|\Lambda\rangle &= C_{1A}|\zeta\rangle_A |\tau\rangle_{14} \\
=& \frac{1}{2} \big[  |0\rangle_A (u|0\rangle+v|1\rangle)_1 |\varepsilon_1\rangle_4
                     +|0\rangle_A (u|0\rangle-v|1\rangle)_1 |\varepsilon_2\rangle_4 \\&
                     +|1\rangle_A (u|0\rangle+v|1\rangle)_1 |\varepsilon_3\rangle_4
                     +|1\rangle_A (u|0\rangle-v|1\rangle)_1 (-|\varepsilon_4\rangle_4)  \big]
\end{aligned}
\end{equation} \par
\subitem{\bf{3.2)}} Now an extra auxiliary qubit $e$ is introduced with initial state $|0\rangle_e$.
Then the controlled-CNOT gate $C_{Ae}$ is applied on system $Ae$ and we obtain the following state
\begin{equation}
\begin{aligned}
|\Omega\rangle
&= C_{Ae}|0\rangle_e |\Lambda\rangle= C_{Ae} C_{1A} |0\rangle_e|\zeta\rangle_A |\tau\rangle_{14} \\
&= \frac{1}{2} \big[  |0\rangle_e |0\rangle_A (u|0\rangle+v|1\rangle)_1 |\varepsilon_1\rangle_4
                     +|0\rangle_e |0\rangle_A (u|0\rangle-v|1\rangle)_1 |\varepsilon_2\rangle_4 \\
&                     +|1\rangle_e |1\rangle_A (u|0\rangle+v|1\rangle)_1 |\varepsilon_3\rangle_4
                     +|1\rangle_e |1\rangle_A (u|0\rangle-v|1\rangle)_1 (-|\varepsilon_4\rangle_4)  \big]
\end{aligned}
\end{equation} \par
\subitem{\bf{3.3)}}To orthogonalize the two nonorthogonal states $u|0\rangle \pm v|1\rangle$, a controlled-U gate $C_{1A}^{U_A}$ is applied on the system $1A$. Here, the $U_A$ is a rotation in a $\pi/2$ around the $\bm{\hat{n}}$ direction, specifically \par
\begin{equation}
U_A = e^{-i \frac{\pi}{2}}e^{i \frac{\pi}{2} \bm{\hat{n}} \sigma^{A}}
\end{equation}
with $\bm{\hat{n}}=\Big( \sqrt{1-\frac{u^2}{v^2}}, 0, \frac{u}{v}  \Big)$ and
$\sigma^{A} = (\sigma^{A}_x, \sigma^{A}_y, \sigma^{A}_z)$.
Another representation of $U_A$ in matrix form is
\begin{equation}
 U_A= \left[
\begin{matrix}
    \frac{u}{v}               &    \sqrt{1-(\frac{u}{v})^2 }   \\
                              &                                \\
    \sqrt{1-(\frac{u}{v})^2}   &     -\frac{u}{v}
\end{matrix}
\right]
\end{equation}
Therefore, we can get the following state
\begin{equation}
\begin{aligned}
|\Gamma\rangle
&= C^{U_A}_{1A}|\Omega\rangle= C^{U_A}_{1A}C_{Ae} C_{1A} |0\rangle_e|\zeta\rangle_A |\tau\rangle_{14} \\
&= \frac{u}{\sqrt{2}} \big[|0\rangle_e |0\rangle_A
                  (|+\rangle_1|\varepsilon_1\rangle_4  + |-\rangle_1|\varepsilon_2\rangle_4)
                   +|1\rangle_e |1\rangle_A
                  (|-\rangle_1|\varepsilon_3\rangle_4  - |+\rangle_1|\varepsilon_4\rangle_4)
                    \big] \\
&+\frac{1}{2}\sqrt{v^2-u^2} \big[
 |0\rangle_e |1\rangle_A |1\rangle_1(|\varepsilon_1\rangle_4-|\varepsilon_2\rangle_4)
+|1\rangle_e |0\rangle_A |1\rangle_1(|\varepsilon_3\rangle_4+|\varepsilon_4\rangle_4) \big]
\end{aligned}
\end{equation} \par
\subitem{\bf{3.4)}}Finally, to reduce the times of measurements,  a controlled-CNOT $C_{Ae}$ is applied on system $Ae$ to get
\begin{equation}
\begin{aligned}
&|\Upsilon\rangle
= C_{Ae}|\Gamma\rangle = C_{Ae}C^{U_A}_{1A}C_{Ae} C_{1A} |0\rangle_e|\zeta\rangle_A |\tau\rangle_{14} \\
&= \frac{u}{\sqrt{2}}|0\rangle_e \big[ |0\rangle_A
                  (|+\rangle_1|\varepsilon_1\rangle_4  + |-\rangle)_1|\varepsilon_2\rangle_4)
                   +|1\rangle_A
                  (|-\rangle_1|\varepsilon_3\rangle_4  - |+\rangle)_1|\varepsilon_4\rangle_4)
                    \big] \\
&+\frac{1}{2}\sqrt{v^2-u^2} |1\rangle_e \big[
  |1\rangle_A |1\rangle_1(|\varepsilon_1\rangle_4-|\varepsilon_2\rangle_4)
+ |0\rangle_A |1\rangle_1(|\varepsilon_3\rangle_4+|\varepsilon_4\rangle_4) \big]
\end{aligned}
\end{equation}
Notice that if $|\tau\rangle = |\tau_1\rangle $, we get
\begin{equation}
\begin{aligned}
|\varepsilon_1\rangle_4-|\varepsilon_2\rangle_4 = (I^{(4)}-\sigma_Z^{(4)} )|\zeta\rangle_4 = 2b|1\rangle, \\
|\varepsilon_3\rangle_4+|\varepsilon_4\rangle_4 = \sigma_X^{(4)} (I^{(4)}+\sigma_Z^{(4)} )|\zeta\rangle_4 = 2a|1\rangle.
\end{aligned}
\end{equation}
Then we have
\begin{equation}
\begin{aligned}
|\Upsilon_1\rangle
&= \frac{u}{\sqrt{2}}|0\rangle_e \big[ |0\rangle_A
                  (|+\rangle_1|\varepsilon_1\rangle_4  + |-\rangle_1|\varepsilon_2\rangle_4)
                   +|1\rangle_A
                  (|-\rangle_1|\varepsilon_3\rangle_4  - |+\rangle_1|\varepsilon_4\rangle_4)
                    \big] \\
& +\sqrt{v^2-u^2} |1\rangle_e |\zeta\rangle_A |1\rangle_1|1\rangle_4
\end{aligned}
\end{equation}
Similarly, if $|\tau\rangle = |\tau_2\rangle $, we get
\begin{equation}
\begin{aligned}
|\varepsilon_1\rangle_4-|\varepsilon_2\rangle_4 = \sigma_X^{(4)} (I^{(4)}- \sigma_Z^{(4)} )|\zeta\rangle_4 = 2b|0\rangle ,  \\
|\varepsilon_3\rangle_4+|\varepsilon_4\rangle_4 =(I^{(4)} + \sigma_Z^{(4)} )|\zeta\rangle_4 = 2a|0\rangle.
\end{aligned}
\end{equation}
Then we have
\begin{equation}
\begin{aligned}
|\Upsilon_2\rangle
&= \frac{u}{\sqrt{2}}|0\rangle_e \big[ |0\rangle_A
                  (|+\rangle_1|\varepsilon_1\rangle_4  + |-\rangle_1|\varepsilon_2\rangle_4)
                   +|1\rangle_A
                  (|-\rangle_1|\varepsilon_3\rangle_4  - |+\rangle_1|\varepsilon_4\rangle_4)
                    \big] \\
& +\sqrt{v^2-u^2} |1\rangle_e |\zeta\rangle_A |1\rangle_1|0\rangle_4
\end{aligned}
\end{equation}

\item{\bf{Step 4 }}
After Alice makes a $Z$ measurement on particle $e$, the following two cases should be considered according to the measurement result. \par

\subitem{\bf{4.1)}} The qubit $e$ is projected onto $|0\rangle_e$ with probability
 \begin{equation}
\begin{aligned}
Pr_{suc}
&= \sum 4 \times \frac{1}{2} |u|^2 \times Pr_i  \\
&= \frac{|\alpha|^2  }{|\alpha|^2+|\eta|^2}[|\alpha|^2+|\eta|^2] \times 2
  + \frac{|\beta|^2  }{|\beta|^2+|\gamma|^2}[|\beta|^2+|\gamma|^2] \times 2 \\
&= 2(|\alpha|^2+|\beta|^2).
\end{aligned}
\end{equation}
 In this case, Alice needs more measurements on qubits ($A$,$1$) in the $Z$-basis and $X$-basis, respectively. The measurement results are able to unambiguously discriminate which state of the qubit $4$ is:
$|\zeta\rangle_4, \sigma_Z^{(4)} |\zeta\rangle_4, \sigma_X^{(4)} |\zeta\rangle_4,$ or $ \sigma_X^{(4)} \sigma_Z^{(4)} |\zeta\rangle_4$. \par

\subitem{\bf{4.2)}}The qubit $e$ is projected onto $|1\rangle_e$ with probability
 \begin{equation}
Pr_{fail}=1-Pr_{suc} = 1 - 2(|\alpha|^2+|\beta|^2) =|\gamma|^2+|\eta|^2 - |\alpha|^2 - |\beta|^2
\end{equation}
and consequently, the unknown state $|\zeta\rangle_A$ is recovered in qubit $A$. Thus for the outcome $|1\rangle_e$ the teleportation fails, but Alice has the intact (un)known state $|\zeta\rangle_A$ again.
\end{description}
Therefore, when the teleportation fails, the whole process can be repeated if the Alice, Bob, and Chika share another quantum channel $|\varphi\rangle_{1234}$.
Denote a variable $X$ be the number of tries required until the first successful teleportation, i.e. $X$ following geometric distribution, then the probability that the $k^{th} (k\ge 1)$ trial is the first success is
\begin{equation}
Pr(X=k) = (1-p)^{k-1}p
\end{equation}
with $p=Pr_{suc}=2(|\alpha|^2+|\beta|^2)$.
Given a $N$, the probability that first $N$ trials would succeed is given by
\begin{equation}
Pr(X \le N) =   \sum_{k=1}^{N}  (1-p)^{k-1}p = 1 - (1-p)^N
\label{eq21}
\end{equation}
\begin{figure}
  \centering
  \begin{minipage}[c]{0.5\textwidth}
  \centering
\includegraphics[height=4cm,width=6cm]{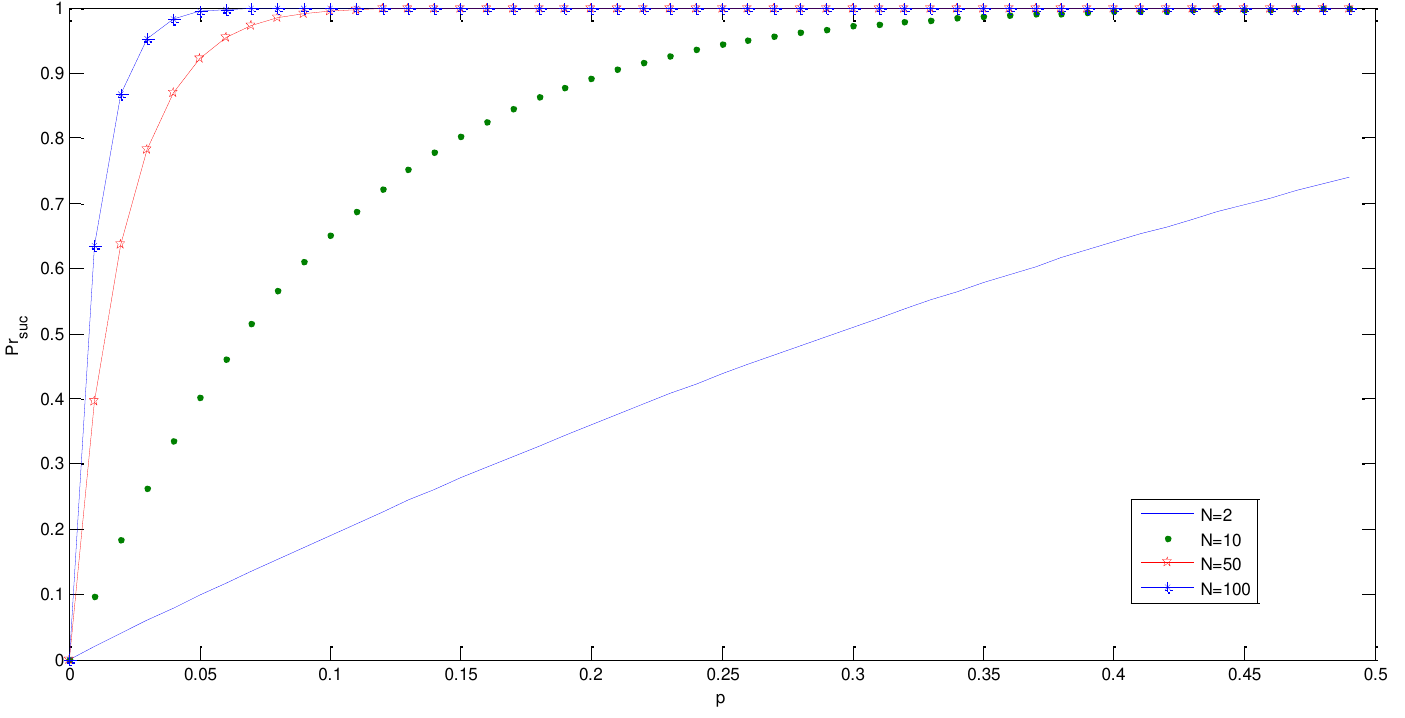}
\end{minipage}%
\begin{minipage}[c]{0.5\textwidth}
\centering
\includegraphics[height=4cm,width=6cm]{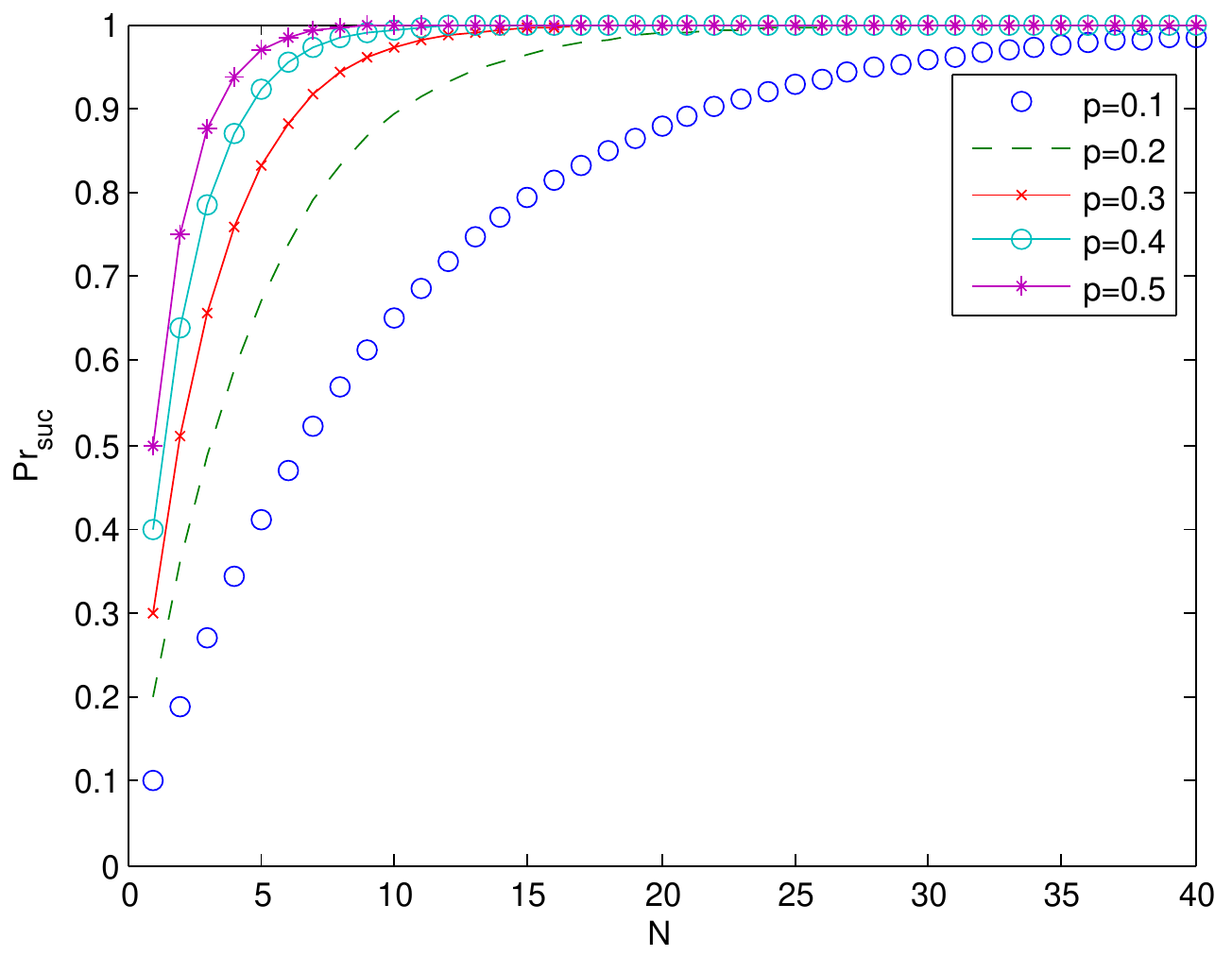}
\end{minipage}
\caption{Total success probability(Eq.~\ref{eq21}) of teleporting the unknown state as the function of $p$ and $N$. (a)The considered $N$'s are 2 (dashed), 10(dotted), 50(dashed-pentagram), and 100(dashed-star).
(b)The considered $p$'s are 0.1(circle), 0.2(dashed-dashed), 0.3(dashed-times), 0.4(dashed-circle), and 0.5(dashed-star).}
\label{fig2}
\end{figure}

\par
Fig.~\ref{fig2} 1(a) gives the variation of $Pr(X\le N)$ as a function of $p$ for different values of $N$, and Figure 1(b) gives the variation of $Pr(X\le N)$ as a function of $N$ for different values of $p$.

\section{Conclusion}
This Comment has pointed out  that  Ram\'irez's scheme\cite{Front. Phys. (2017) 12: 120306} can not realize repeatedly teleportation without another copy of $|\zeta\rangle$ while teleportation process fails. The main reason is that once Alice performs Bell measurement on her only qubit pair $(A,1)$, all the information of  $|\zeta\rangle$ is shifted to Chika. To overcome the problem, Alice performs several unitary operators and $\sigma_Z$ measurements instead of Bell measurements, and Alice measurement results decide whether Chika can recover $|\zeta\rangle$ or not. The analysis of our proposed protocol shows that the $|\zeta\rangle$ stays still in qubit $A$ if the teleportation process fails, so the teleportation process can be repeated using another quantum channel $|\varphi\rangle_{1234}$. Finally, the simulation illustrates that the success probability of te
leporting the unknown state quickly approximates to 1 with the increasing of $p$ when $N$ is large ($N \ge 50$) or with the increasing of $N$ when $p$ is large ($p \ge 0.3$).




\end{document}